\documentclass[graybox]{svmult}

\usepackage{mathptmx}       
\usepackage{helvet}         
\usepackage{courier}        
\usepackage{type1cm}        
\usepackage{makeidx}         
\usepackage{graphicx}        
\usepackage{multicol}        
\usepackage[bottom]{footmisc}
\usepackage{amsmath,amssymb}

\def\NN{\mathbb{N}}
\def\ZZ{\mathbb{Z}}
\def\RR{\mathbb{R}}

 \def\wh{\widehat}

\def\wick#1{{\colon\!#1\colon\!}} 

\def\bea{\begin{eqnarray}} \def\eea{\end{eqnarray}}
\def\eref#1{Eq.\ (\ref{#1})}

\renewcommand{\medskip}{\smallskip}

\begin{document}

\title*{New methods in conformal partial wave analysis}

\titlerunning{Partial wave analysis} 

\author{Christoph Neumann,
Karl-Henning Rehren and Lena Wallenhorst
}

\institute{Christoph Neumann \at Institut f\"ur Theoretische Physik, Universit\"at G\"ottingen, 
Friedrich-Hund-Platz 1, D-37077 G\"ottingen, Germany, {e-mail: christoph.neumann@theorie.physik.uni-goettingen.de}
\and Karl-Henning Rehren \at Institut f\"ur Theoretische Physik, Universit\"at G\"ottingen, 
Friedrich-Hund-Platz 1, D-37077 G\"ottingen, Germany; Courant Research Centre ``Higher Order Structures in
  Mathematics'', Universit\"at G\"ottingen, Bunsenstr. 3--5, D-37073
  G\"ottingen, {email: rehren@theorie.physik.uni-goettingen.de}
\and Lena Wallenhorst \at Institut f\"ur Theoretische Physik, Universit\"at G\"ottingen, 
Friedrich-Hund-Platz 1, D-37077 G\"ottingen, Germany,
{email: lena.wallenhorst@theorie.physik.uni-goettingen.de}
}

\maketitle

\vskip-5mm

\abstract{We report on progress concerning the partial wave analysis
of higher correlation functions in conformal quantum field theory.
\footnote{To appear in the proceedings of the conference ``LT-9 -- Lie
  Theory and Applications in Physics'', Varna, Bulgaria, June 2011}}

\renewcommand{\theequation}{\thesection.\arabic{equation}}

\section{Introduction}
\label{sec:Intro}

Partial wave analysis (PWA) is a powerful tool in conformal quantum field
theory. It gives not only information about the field content and the
operator product expansion (OPE) of a model \cite{M76,LR}, but can
also be used for probing the positivity of the inner product induced by the
correlation functions (Wightman positivity) \cite{NRT05}.  

Positivity is difficult to establish because it is a nonlinear
property. It also necessarily involves correlation functions of any 
number of fields \cite{Y}. The most prominent example is the
classification of central charges below 1 of the Virasoro algebra.  
An example in four spacetime dimensions (4D) is the result that
conformal scalar fields with global conformal invariance (GCI,
\cite{NT01}) are necessarily Wick squares of free fields \cite{NRT06}, 
and cannot couple in a nontrivial manner to other fields \cite{BNRT}.

\medskip

While conformal PWA for 4-point functions is well understood
\cite{DO}, we intend to develop methods for higher
correlation functions. The basic task is to decompose a correlation function 
of conformally covariant fields into a sum over partial waves 
\bea
\label{PWE}(\Omega,\phi_1(x_1)\dots\phi_n(x_n)\Omega)
=\sum_{\lambda}(\Omega,\phi_1(x_1)\dots\phi_{k-1}(x_{k-1})
\Pi_{\lambda}\phi_{k}(x_{k})\cdots\phi_n(x_n)\Omega), \qquad
\eea
where $\Pi_{\lambda}$ is the projection to the subspace of
the Hilbert space which carries the irreducible representation
$\lambda$ of the conformal group. A projection can be
inserted in any position within the correlation, so that the $n$-point
partial waves depend on $n-1$ representations, where the first and
last projections are redundant because they are fixed by the first and
the last field. 

In principle, the non-vanishing partial waves give information about
the contributions to the OPE of two or more fields \cite{M76}. Since a
projection is a positive operator, each partial wave contribution of
the form 
$$(\Omega, \phi'(x_1)\dots\phi(x_n)\Pi_{\lambda}\phi(x_{n+1})\dots\phi'(x_{2n})\Omega)$$
must separately satisfy Wightman positivity (i.e., after smearing with
test functions $\overline{f(x_n,\dots,x_1)}f(x_{n+1},\dots,x_{2n})$ it must yield a 
non-negative number which is the norm square of the vector 
$\Pi(\phi\otimes\dots\otimes\phi')(f)\Omega$). More generally, partial
waves are subject to Cauchy-Schwartz type inequalities. 

Now, partial waves are to a large extent determined by
conformal symmetry, being solutions to eigenvalue equations for the 
Casimir operators of the conformal group. Therefore, the positivity
requirement reduces to the positivity of a numerical coefficient, the
partial wave amplitude, which multiplies a model-independent partial
wave function \cite{NRT05}. 

\medskip

Conformal PWA is by now mostly limited to 4-point functions, because the
higher partial waves are not sufficiently well known. Even for 4
points, the determination of partial waves in 4D required a
considerable effort \cite{DO}. Moreover, the decomposition of a given
correlation function into a known system of partial waves may not be a
straight-forward task without a suitable notion of orthogonality
between the partial waves. Some progress was made in \cite{NRT05}
giving a systematic expansion formula for scalar 4-point partial
waves, and in \cite{R10} for a suitable notion of orthogonality.  

\medskip

In this note, we report some further intermediate progress. In Sect.\
\ref{2Dps}, we present a power series representation \eref{pwexp} for
general $n$-point partial waves in two spacetime dimensions (2D) for all $n$,
extending known formulae for $n\leq 4$. In 4D, however, such an expansion seems
unrealistic because of the complicated structure of the higher-order
Casimir operators which the partial waves must diagonalize, and
because the partial waves are no longer unique. 

In Sect.\ \ref{4Dint} we therefore present an alternative to the actual
decomposition \eref{PWE}, which is applicable also in 4D. The idea is
a successive reduction of $n$-point functions to $n-1$-functions,
in terms of local linear maps $\phi_1(x_1)\phi_2(x_2)\Omega \to
\phi_\lambda(x)\Omega$ selecting each contribution to the OPE of the
last two (or the first two) fields in the correlation. Our main result
is the characterization of these linear maps as partial differential
operators that intertwine the respective representations of the conformal
group. This property is encoded in \eref{itwS}, which is subsequently
solved. Acting on the correlation functions, the intertwiners effectuate the
desired reduction. As we shall see, this method is applicable only for
representations of integer scaling dimension (otherwise, the
differential operators would have to be replaced by integral kernels
\cite{DMPPT}, and locality would become a nontrivial issue). 

This method is therefore well-suited for QFT with global conformal
invariance, where all correlation functions are
rational functions \cite{NT01}. We shall apply it in Sect.\ \ref{Exo}
to address the problem of positivity of a class of ``exotic'' higher
($n\geq6$) correlation structures of twist 2. The motivation is the following. 

Twist-2 contributions in free field theories above the unitarity bound
arise from quadratic Wick products such as
$\wick{\varphi^*(x_1)\varphi(x_2)}$ or
$x_{12\mu}\wick{\overline\psi(x_1)\gamma^\mu\psi(x_2)}$, in which   
each factor can be contracted ``only once'', so that both variables
can only have poles w.r.t.\ one other variable. 
In contrast, the exotic structures contain so-called double
poles, thus indicating
a nontrivial theory. These are strongly constrained by the
conservation laws for twist-2 fields \cite{NRT06}, allowing for a
classification \cite{B09}. In particular, they cannot arise in
correlations of less than six fields. While the exotic structures
satisfy all linear properties, it remains an open problem whether they
are compatible with positivity.  

First steps of the positivity analysis of the simplest exotic
structure will be reported in Sect.\ \ref{Exo}.

\section{Higher chiral partial waves}
\label{2Dps}
\setcounter{equation}{0}

Irreducible representations $\lambda$ of the conformal group are
eigenspaces of the Casimir operators. Thus, correlation functions with
projections onto irreducible subrepresentations inserted:
\bea\label{pw}
\langle \Omega,\phi_1(x_1) \Pi_{\lambda_1}\phi_2(x_2) \cdots
\Pi_{\lambda_{i-1}}\phi_i(x_i)
\Pi_{\lambda_i}\cdots\phi_{n-1}(x_{n-1}) \Pi_{\lambda_{n-1}}\phi_n(x_n)
\Omega \rangle 
\qquad\eea
are eigenvectors of the corresponding differential operators arising
by commuting the conformal generators with the fields. Partial waves
are, by definition, solutions to the same eigenvalue differential
equations, with some standard normalization. These are ``universal'' in
the sense that they are completely determined by conformal
symmetry. They depend on the sequence of representations
$(\mu_i)_{i=1\dots n}$ of the fields $\phi_i$ in \eref{pw}, and on the 
sequence of representations $(\lambda_i)_{i=1\dots n-1}$ of the
projections, where $\lambda_1=\overline\mu_1$ and $\lambda_{n-1}=\mu_n$ are redundant. 

\medskip

The projected
correlations \eref{pw} are multiples of the partial waves. The coefficients
contain model-specific information, and Wightman positivity can be
formulated as a system of numerical inequalities on the partial wave coefficients \cite{NRT05}. 

The conformal Lie algebra in 4D, $so(4,2)$, has three Casimir operators
(quadratic, cubic and quartic in the generators). In contrast, the
conformal Lie algebra in 2D factorizes: $so(2,2)\sim sl(2,\RR)\oplus
sl(2,\RR)$, and each $sl(2,\RR)$ has one quadratic Casimir
operator. For this reason, the Casimir eigenvalue differential
equations are much simpler (both, to write down and to solve) in
2D.

The relevant positive-energy representations of $sl(2,\RR)\oplus
sl(2,\RR)$ are parameterized by the chiral scaling dimensions $d_\pm$,
such that $d_++d_-$ is the total scaling dimension, and $d_+-d_-$ the
helicity.  

Because of the chiral factorization of the conformal group, also the
partial waves factorize. In the sequel, we display only chiral partial
waves as functions of either $x_+=t+x$ or $x_-=t-x$, and suppress the
subscript. Thus, a general projected correlation function in 2D has
the form of a product of two chiral functions
\bea\label{pw2D}
\langle \Omega, \phi_1(x_1) \Pi_{a_1}\phi_2(x_2) \cdots \Pi_{a_{i-1}}\phi_i(x_i)
\Pi_{a_i}\cdots\phi_{n-1}(x_{n-1}) \Pi_{a_{n-1}}\phi_n(x_n) \Omega\rangle
\qquad\eea
where the chiral fields $\phi_i$ have chiral dimensions
$d_i$, and $\Pi_a$ are the projections onto the chiral representations
with chiral scaling dimension $a$. In particular, $a_1=d_1$ and
$a_{n-1}=d_n$ are fixed.  

The Casimir eigenvalue equation for the projector insertion $\Pi_{a_i}$ reads
\bea\notag \Big(\sum_{i<j<k} x_{jk}^2\partial_j\partial_k
+2\sum_{i<j,k} d_j (x_{jk}\partial_k)  + \sum_{i < k} d_k -
\Big(\sum_{i< k} d_k\Big)^2 \Big)
\langle\dots\Pi_{a_i}\phi_{i+1}(x_{i+1})\dots\rangle \\ = (a_i - a_i^2) \langle\dots\Pi_{a_i}\phi_{i+1}(x_{i+1})\dots\rangle,
\notag \eea
which is equivalent by conformal invariance to 
\bea\notag \Big(\sum_{j<k\leq i} x_{jk}^2\partial_j\partial_k
+2\sum_{j,k\leq i} d_j (x_{jk}\partial_k)  + \sum_{ k\leq i} d_k -
\Big(\sum_{k\leq i} d_k\Big)^2 \Big)
\langle\dots\phi_{i}(x_{i})\Pi_{a_i}\dots\rangle \\ = (a_i - a_i^2) \langle\dots\phi_{i}(x_{i})\Pi_{a_i}\dots\rangle.
\notag\eea
In principle, these equations can be reformulated in terms of $n-3$
independent conformal cross ratios. It turns out convenient to choose
$$u_k= \frac{x_{kk+1}x_{k+2k+3}}{x_{kk+2}x_{k+1k+3}}.$$
We have worked out the invariant differential equations for $n\leq
6$ points: Let
$$(\ref{pw2D}) =\frac{f(u_1,u_2,u_3)}
{x_{12}^{d_1+d_2-d_3}x_{13}^{d_1+d_3-d_2}x_{23}^{d_2+d_3-d_1}\cdot
x_{45}^{d_4+d_5-d_6}x_{46}^{d_4+d_6-d_5}x_{56}^{d_5+d_6-d_4}}.$$
Then (with the Euler operators $E_i=u_i\partial_{u_i}$) 
\bea\notag\label{invcas}
(E_1+d_3-a_2)(E_1+d_3+a_2-1) f &=&
u_1(E_1+E_2)(E_1+d_1-d_2+d_3)f, 
\\ \notag 
(E_2-a_3)(E_2+a_3-1) f &=& u_2(E_2+E_1)(E_2+E_3)f, \qquad \\
(E_3+d_4-a_4)(E_3+d_4+a_4-1) f &=&
u_3(E_3+E_2)(E_3+d_6-d_5+d_4)f.\qquad\qquad
\eea
(The cases $n<6$ are covered by admitting the trivial field $\mathbf
1$ of dimension $0$.) This system is obviously symmetric under
hermitean conjugation $1,2,\dots,6\to 6,5,\dots,1$. It can be
recursively solved as a power series with leading powers
$u_1^{a_2-d_3}u_2^{a_3}u_3^{a_4-d_4}$.  

\medskip

From the solution with $n\leq 6$, we have extrapolated the general
power series expansion for all $n$, as follows. By default, we
put $a_0=a_n:=0$, and $\ell_0=\ell_{n-2}:=0$. 

\medskip

{\bf Proposition 1:} {\it The general chiral $n$-point partial wave is} 
\bea\label{pwexp}
\sum_{\ell_1,\dots,\ell_{n-3}\geq0}\frac{\prod_{j=1}^{n-2}x_{jj+2}^{d_{j+1}-a_j-a_{j+1}}(a_j+a_{j+1}-d_{j+1})_{\ell_{j-1}+\ell_j}}{\prod_{i=1}^{n-1}x_{ii+1}^{d_i+d_{i+1}-a_{i-1}-a_{i+1}}}
 \cdot \prod_{k=1}^{n-3}
\frac{u_k^{\ell_k}}{\ell_k!(2a_{k+1})_{\ell_k}} .
\qquad \eea
This formula has a remarkable ``short-range'' feature: It involves
only coordinate distances $x_{ij}$ with $j=i+1$ or $i+2$. The
powers of $x_{ii+1}$ and $x_{ii+2}$ depend only on the dimensions of
the fields $\phi_i$, $\phi_{i+1}$, respectively $\phi_{i+1}$, and their 
adjacent projections, apart from the summation indices $\ell$. The
same is true for the numerical coefficients. 

For $n=3$ points, this is just the 3-point function. For $n=4,5,6$
points, we have derived this formula by solving the differential
equations \eref{invcas} for the Casimir eigenvalues. For $n=4$, the sum is a
hypergeometric series, and \eref{pwexp} coincides with well-known formulas. 

\medskip

One way to prove the \eref{pwexp} for all $n$ is an
application of the method discussed in the next section. There, we
introduce ``intertwining'' differential operators $\iota\circ\wh E_h$
with the distinguishing property that they annihilate all partial
waves carrying the ``wrong'' representation $a\neq h$, and reduce the
$n$-point partial wave carrying the representation $a=h$ to an
$(n-1)$-point partial wave with the first pair of fields replaced by
$\phi_0$ of dimension $h$.  

Therefore, it is sufficient to show that this is true for our ``candidate''
partial waves \eref{pwexp}. With \eref{chiral-d}, we have to apply the
differential operator 
$$\wh E_h\equiv E_h\circ x_{12}^{d_1+d_2} = 
\Big(\sum_{p+q=h}\frac{(q-b)_p}{p!}\frac{(p+b)_q}{q!} \;\partial_1^p(-\partial_2)^q\Big)
\circ x_{12}^{d_1+d_2},$$
where $b=d_1-d_2$, to \eref{pwexp}, and then equate $x_1=x_2$. The
result must be $\delta_{ha_2}$ times the reduced partial wave. 

To do this, we have to exhibit all terms that involve
$x_1$ or $x_2$. \eref{pwexp} can be arranged as $x_{12}^{-d_1-d_2}$
times the sum
$\sum_{\ell_2,\dots,\ell_{n-3}}$ over 
\bea\label{subsum}
 \left(\frac{x_{12}}{x_{13}x_{24}}\right)^{a}\left(\frac{x_{23}}{x_{13}}\right)^{b}\left(\frac{x_{23}}{x_{24}}\right)^{c}\sum_{\ell\geq0}
\frac{(a+b)_\ell(a+c)_\ell u^\ell}{\ell!(2a)_\ell}\times 
\hbox{remaining factors},
\qquad \eea
where $a\equiv a_1$, $b\equiv d_1-d_2$, $c\equiv a_3-d_3+\ell_2$. Notice that for each $\ell_2$, the
sum over $\ell$ is a 4-point partial wave where the $4^{\rm th}$ field
has dimension $a_3+\ell_3$. Thus, knowing that \eref{pwexp} correctly
reproduces the 4-point partial waves, and that $\iota\circ \wh E_h$ reduces
4-point partial waves to 3-point partial waves, the same must be true
for the higher partial waves. 

However, we have not been able to evaluate the result of
$E_h^{d_1,d_2}$ on the power series \eref{subsum}, and verify this
indirect conclusion by a direct computation. Only for $n=3$ this can
be done by the following argument. For $n=3$, one has $c=0$ in
\eref{subsum}, only $\ell=0$ contributes, and there are no ``remaining
factors''. Then

(i) Because $E_h^{d_1,d_2}$ is a differential operator of order $h$, it
annihilates the 3-point function whenever $h<a$, due to the surviving
factors of $x_{12}$. 

(ii) Writing $\frac{x_{12}}{x_{13}x_{23}}=\frac 1{x_{23}}-\frac
1{x_{13}}$ and performing the binomial expansion of its powers,
$E_h^{d_1,d_2}$ can easily be applied. It is then seen by inspection
that the resulting series is symmetric under the exchange $a\leftrightarrow
h$. Therefore, it also vanishes whenever $h>a$.

(iii) When $h=a$, all derivatives must hit the factor $x_{12}^a$. That
the result is the 2-point function, is then obvious. 

For $n>3$, the Leibniz rule produces multiples sums which are not
easy to handle. But a trick helps: The 
sum in \eref{subsum} equals ${}_2F_1(a+b,a+c;2a;u)$. 
We then use the identity
\bea\notag 
\frac{x_{34}^{2a-1}}{(x_{13}x_{24})^a}\left(\frac{x_{23}}{x_{13}}\right)^{b}\left(\frac{x_{23}}{x_{24}}\right)^{c}\cdot
{}_2F_1(a+b,a+c;2a;u) =  \qquad\qquad\qquad \\ = \frac{\Gamma(2a)}{\Gamma(a+c)\Gamma(a-c)}\int_{x_3}^{x_4}dx \,
(x_1-x)^{-a-b}(x_2-x)^{-a+b}(x_3-x)^{a+c-1}(x-x_4)^{a-c-1},\notag\eea
which can be established by direct
computation: namely, the change of variables
$t=\frac{x_{24}(x_3-x)}{x_{34}(x_2-x)}$ yields precisely the standard
integral representation of the hypergeometric function.

Therefore, each term 
\eref{subsum} is, as far as its dependence on $x_1$ and $x_2$ is concerned,
an integral over a 3-point function. Thus, we only have to evaluate 
$E_h^{d_1,d_2}$ on a 3-point function, which can be done as
before. The remaining integral is again of the hypergeometric type
(after the change of variables $t=-\frac{(x_3-x)}{x_{34}}$), and
reproduces precisely the necessary ``leading'' factors for the
$(n-1)$-point partial wave \eref{pwexp}.

\medskip

From this, we conclude that \eref{pwexp} indeed is the correct power series
expansion of general $n$-point chiral partial waves.

\section{Intertwining differential operators}
\label{4Dint}
\setcounter{equation}{0}

Let $\phi_1$ and $\phi_2$ be two conformal fields transforming in
representations $\mu_1$ and $\mu_2$. We shall determine differential
operators $\wh E_\lambda$ w.r.t.\ $x_1$ and $x_2$ 
such that 
\bea\label{reduction}
\phi_\lambda(x):=\iota_x\circ \wh E_\lambda\, \phi_1(x_1)\phi_2(x_2)
\eea
transforms like a conformal field in the representation
$\lambda$. Here, $\iota_x$ is the evaluation map $\iota_x (f)
= f(x_1,x_2)\vert_{x_1=x_2=x} = f(x,x)$.

It will become clear below that such operators exist only
when the scaling dimensions satisfy $d_\lambda-d_1-d_2\in\ZZ$. They
can therefore be expected to be exhaustive (w.r.t.\ $\lambda$) only in
a globally conformal invariant (GCI) theory.   

Such operators have been presented previously \cite[Sect.\ VI.B]{MPT}
for the special case of $\phi_1$ and $\phi_2$ being two (complex
conjugate) canonical scalar massless Klein-Gordon fields of dimension
$1$, in order to extract the current, the stress-energy tensor and
higher conserved symmetric traceless tensor fields from
$\wick{\phi^*\phi}$. The same operators actually can be used also for
scalar biharmonic bifields $V(x_1,x_2)$ which collect the twist-2
contribution in any product of two scalar fields of equal dimension
\cite{NRT06}, where biharmonicity, i.e., the wave equation w.r.t.\
both arguments is exploited in an essential way. 
We shall reproduce these operators, but there will be
additional terms including the wave operators, so that
\eref{reduction} is true without using the equation of motion, or
biharmonicity.  

\medskip

By conformal covariance, the assumed transformation behaviour of 
\eref{reduction} implies
$$\iota_x\circ \wh E_\lambda
(\phi_\mu(y)\Omega,\phi_1(x_1)\phi_2(x_2)\Omega) = \delta_{\lambda\mu}
(\phi_\mu(y)\Omega,\phi_\lambda(x)\Omega),$$
i.e., the operator annihilates all 3-point functions with fields in
the ``wrong'' representation. In particular, if applied to the vacuum
operator product expansion \cite{M76}
$$\phi_1(x_1)\phi_2(x_2)\Omega = \sum_\mu \int dx \,
K_\mu^{\mu_1\mu_2}(x_1,x_2;x)\phi_\mu(x)\Omega \, ,$$ 
where $K_\mu^{\mu_1\mu_2}$ are certain integral kernels, it will
annihilate all contributions $\mu\neq\lambda$, and if applied 
to a correlation function, it will annihilate all partial waves with
$\mu\neq\lambda$ in the $1$-$2$-channel, and reduce the contribution
with $\mu=\lambda$ to an $n-1$-point partial wave. Thanks to the
latter feature, one can perform a partial wave analysis without
actually knowing the partial waves, cf.\ Sect.\ \ref{Exo}.

\medskip

Let us now proceed to determine the differential operators. 

For definiteness, we specialize to $\mu_1=\mu_2$ to be scalar
representations of dimension $d_1=d_2=d$. In this case, only symmetric
traceless tensor representations $\lambda$ can occur \cite{M76}. 
It is convenient to write $\lambda=(\kappa,L)$ where $L$ is the tensor
rank, and $2\kappa$ the ``twist'', such that the scaling dimension is
$d=2\kappa+L$. The unitarity bound requires $\kappa\geq 0$ for $L=0$, and
$\kappa\geq1$ for $L>0$. We write a symmetric traceless tensor as
$T(v)=T^{\mu_1\dots \mu_L}v_{\mu_1}\dots v_{\mu_1}$ which is a
homogeneous polynomial of degree $L$ in the polarization vector
$v$. Tracelessness is equivalent to the harmonic equation 
$\square_vT(v)=0$.  \eref{reduction}
implies that $\wh E_{\kappa L}$ is a harmonic homogeneous polynomial of
degree $L$ in the polarization vector $v$. The harmonic part of any polynomial in $v$ is
uniquely determined \cite{BT}, so it is sufficient to know $\wh E_{\kappa L}$ up
to terms involving $v^2$.

\medskip 

Let $T=P_\mu,D$, $M_{\mu\nu},K_\mu$ be the generators of translations,
dilations, Lorentz and special conformal transformations,
respectively, and  
$$i[T,\phi(x)]=t^\lambda_x\;\phi(x)$$
the commutation relations with covariant (``quasiprimary'') fields,
where $t^{\kappa L}=\partial$ for the translations,
$=(x\partial+d_\lambda)$ for the scale transformations,
$=x\wedge\partial + v\wedge\partial_v$ for the Lorentz
transformations, and $=2x(x\partial)-x^2\partial + 2(v(x\partial_v)-
(xv)\partial_v) + 2d_\lambda x$ for the special conformal
transformations. For the tensor representations,
$d_\lambda=2\kappa+L$, while for the scalar representations
$\mu_1=\mu_2$ the $v$-terms are absent and $d_\mu=d$. 

\newpage

Commuting the generators with \eref{reduction}, the assumption that
$\phi_\lambda$ transforms in the representation $\lambda$ is
equivalent to the intertwining relations 
\bea\label{itw1}
\iota\circ
\wh E_{\lambda}\circ\big(t^{\mu_1}_{x_1}+t^{\mu_2}_{x_2}\big)
=t^\lambda_{x}\circ\iota\circ \wh E_{\lambda}.
\eea
In the case at hand, we make an ansatz
\bea \label{ans}
\wh E_\lambda = E_{\kappa L}(x_i,\partial_i,v) \circ (x_{12}^2)^d\,.
\eea
Notice that by virtue of the pole bounds \cite{NT01}, any correlation
function of $\phi_1(x_1)\phi_2(x_2)$ is not more singular than
$(x_{12}^2)^{-d}$, so that the differential operators $E_{\kappa L}$
act on a regular function, and the subsequent evaluation $\iota_x$ is
possible (provided $E_{\kappa L}$ is regular).

\medskip

Next, we evaluate the intertwining relations \eref{itw1}. 
They tell us in turn: 

Translations: $(\partial_1+\partial_2)E_{\kappa L}=0$. Thus the
differential operators do not involve the coordinate $x_1+x_2$. Since
$E_{\kappa L}$ is followed by the evaluation map $\iota_x$, we may
also assume that it does not involve the difference coordinate
$x_1-x_2$, hence $E_{\kappa L}$ involves only derivatives and the
polarization vector $v$. Let us denote by $\nabla_i$ the derivatives
with respect to the ``variables'' $\partial_i$ of $E_{\kappa
  L}(\partial_1,\partial_2,v)$. 

Scale transformations:
$(\partial_1\nabla_1+\partial_2\nabla_2)E_{\kappa
  L}=(2\kappa+L)E_{\kappa L}$. Thus, $E_{\kappa L}$ is homogeneous of
degree $2\kappa+L$ in the derivatives $\partial_i$. 

Lorentz transformations: $(\partial_1\wedge\nabla_1
+ \partial_2\wedge\nabla_2 + v\wedge \partial_v)E_{\kappa L}=0$. Thus,
$E_{\kappa L}$ is a Lorentz scalar. It is therefore a function of
$(\partial_i\partial_j)$, $(v\partial_i)$ and $v^2$. Together with the
known homogeneities in $v$ and in $\partial_i$, it can be a polynomial
in the derivatives only if $\kappa$ is an integer. This is in perfect
agreement with GCI because tensor-scalar-scalar 3-point functions are
rational only if the twist $2\kappa$ is even. 

Special conformal transformations: While the previous intertwining
conditions gave information about the gross structure of $E_{\kappa
L}$, the special conformal transformations yield a 
differential equation that specifies the operators completely.  

\medskip

{\bf Proposition 2:} \textit{Given the previous specifications of
  $E_{\kappa L}(\partial_1,\partial_2,v)$ in \eref{ans} as homogeneous polynomials
  (of degrees depending on the parameters $\kappa$ and 
  $L$), the intertwining condition \eref{itw1} is equivalent to }
\bea\label{itwS}
\Big(2(\partial_1\nabla_1)\nabla_1-\partial_1\nabla_1^2 + 2(\partial_2\nabla_2)\nabla_2-\partial_2\nabla_2^2 
\Big)
E_{\kappa L}(\partial_1,\partial_2,v)=0.
\eea

One may directly solve these equations with a polynomial ansatz for
$E_{\kappa L}$ with the specified homogeneities. A more systematic way
is to write 
\bea\label{red}
E_{\kappa
  L}(\partial_1,\partial_2,v)=(\partial_1\partial_2)^\kappa\cdot \big[\big((v\partial_1)+(v\partial_2)\big)^L
\cdot e_{\kappa L}(p,q,r)\big]_0
\eea
where $p=\frac{\partial_1^2}{(\partial_1\partial_2)}$, $q=\frac{\partial_2^2}{(\partial_1\partial_2)}$, and 
$r=\frac{(v\partial_1)-(v\partial_2)}{(v\partial_1)+(v\partial_2)}$. Clearly, $e_{\kappa L}$ must
be a polynomial of degree at most $L$ in $r$, and degree of at most
$\kappa$ in $p$ and $q$. The notation $[P(v)]_0$ stands for the harmonic
part of the polynomial $P(v)$. The variable $v^2$ does not
appear explicitly, because the harmonic part $[v^2 Q(v)]_0=0$ for any
polynomial $Q$ \cite{BT}.

\newpage

With this ansatz, the differential equation \eref{itwS} turns into 
the system of three PDE for $e_{\kappa L}(p,q,r)$: 
\bea\label{system1}
\big(L(L-1)+(1-r^2)\partial_r^2+2\kappa(L-r\partial_r) +
2(p\partial_p-q\partial_q)\partial_r\big)\;e_{\kappa L}=0,
\qquad\\[1mm]
\notag
\big[4\big(p\partial_p-1\big)\partial_p-q(\kappa-p\partial_p-q\partial_q)(\kappa-1-p\partial_p-q\partial_q)
+\qquad\qquad\qquad \\ 
+
2(\kappa-p\partial_p-q\partial_q)\big(\kappa-1-p\partial_p+q\partial_q+(r-1)\partial_r\big) \big]\;e_{\kappa L}=0,
\label{system2}
\qquad\\[1mm] \notag
\big[4\big(q\partial_q-1\big)\partial_q-p(\kappa-p\partial_p-q\partial_q)(\kappa-1-p\partial_p-q\partial_q)
+\qquad\qquad\qquad \\ 
+
2(\kappa-p\partial_p-q\partial_q)\big(\kappa-1+p\partial_p-q\partial_q+(r+1)\partial_r\big) \big]\;e_{\kappa L}=0.
\label{system3}
\qquad\eea

One may repeat the same strategy in 2D. In this
case, the intertwining operators factorize into two chiral operators,
labelled by the chiral dimensions $h_\pm$. These are polynomial
functions in the chiral (one-dimensional) partial derivatives
$\partial_1$ and $\partial_2$. Following the same line of arguments as
in 4D, one finds the chiral intertwining condition 
$$\big(\partial_1\nabla_1^2+\partial_2\nabla_2^2\big)E_{h}(\partial_1,\partial_2)=0,$$
where $E_h(\partial_1,\partial_2)$ is a homogeneous polynomial of degree $h$. Writing
$E_h=(\partial_1+\partial_2)^h\cdot
e_h\big(\frac{\partial_1-\partial_2}{\partial_1+\partial_2}\big)$,
this reduces to the differential equation for $e_h(r)$
$$\big(h(h-1)+(1-r^2)\partial_r^2\big)e_h(r)=0,$$ 
which is exactly the same as the case $\kappa=0$, $L=h$ of \eref{system1}.

\medskip

Notice that in 4D, 
representations $(0,L)$ with $L\neq0$ are below the unitarity
bound. Such representations must not contribute to a correlation
function. Thus, any admissible correlation function must be
annihilated by the operators $\iota\circ \wh E_{0L}$.  
The solution for $\kappa=0$ is
$$e_{0L}(r) = (1-r^2)\partial_r P_{L-1}(r),$$
where $P_n$ are the Legendre polynomials. Using \eref{red}, this gives
\bea\label{K=0}
E_{0L}(\partial_1,\partial_2,v) =
\sum_{p+q=L}
 \frac{(q)_p}{p!}\frac{(p)_q}{q!} \, \Big[(v\partial_1)^{p}(-v\partial_2)^{q}
\Big]_0\;,
\eea 
or (in the chiral case)
\bea\label{chiral}
E_h(\partial_1,\partial_2) =
\sum_{p+q=h}
 \frac{(q)_p}{p!}\frac{(p)_q}{q!} \, \partial_1^{p}(-\partial_2)^{q}
\,.\eea

For $\kappa>0$, we may expand
$$e_{\kappa L}(p,q,r) = \sum_{m,n\geq0, m+n\leq\kappa} p^mq^n
e_{\kappa L;mn}(r).$$
Then \eref{system1} must hold for each term $p^mq^ne_{\kappa L;mn}(r)$
separately, giving  
\bea\label{systemmn}
\big((1-r^2)\partial_r^2 -2\kappa r\partial_r + 2(m-n)\partial_r + L(L+2\kappa-1)\big)e_{\kappa L;mn}(r) =0.
\qquad\eea
This equation involves only the difference $m-n =: \delta$. It is
solved by  polynomials of degree $L$ with the symmetry $f_{\kappa
  L;\delta}(r)=(-1)^L f_{\kappa L;-\delta}(-r)$:
\bea\label{fdelta} 
f_{\kappa L;\delta}(r)=
(\kappa-\delta)_L\cdot{}_2F_1\Big(-L,L+2\kappa-1;\kappa-\delta;\frac{1-r}2\Big) .\qquad
\eea
Thus, to solve \eref{systemmn} it remains to determine only the
coefficients in 
\bea \label{ece}
e_{\kappa L;mn}(r) = c_{\kappa L;mn}
\cdot f_{\kappa L;m-n}(r) .
\eea
Indeed, the remaining \eref{system2} and \eref{system3}
turn into the recursive system  
\bea \label{rec1}\notag
4(m^2-1)c_{\kappa L;m+1,n} +2(\kappa-m-n)(L+\kappa-1-m+n) c_{\kappa L;m,n}
\\ -(\kappa-m-n)(\kappa-m-n+1) c_{\kappa L;m,n-1} =0 , \\
\label{rec2}\notag
4(n^2-1)c_{\kappa L;m,n+1} +2(\kappa-m-n)(L+\kappa-1+m-n) c_{\kappa L;m,n}
\\ -(\kappa-m-n)(\kappa-m-n+1) c_{\kappa L;m-1,n} =0 . \eea
Here, we have used the fact (\cite[Eqs.\ 15.2.14 and 15.2.16]{AS}) that the
differential operators 
$$ A^\pm_{\kappa L,\delta} := \frac{(r\mp 1)
\partial_r+\kappa-1\mp\delta}{L+\kappa-1\mp\delta}$$
act as raising and lowering operators for the parameter $\delta$:
\bea\label{Apm}A^\pm_{\kappa L,\delta} \; f_{\kappa L;\delta} =
f_{\kappa L;\delta\pm 1}.
\eea 
We conclude:

\medskip

{\bf Proposition 3:} \textit{The
intertwining differential operators in \eref{reduction} are given by 
$$\widehat E_{\kappa L} = \!\sum_{m+n\leq\kappa}\!\!c_{\kappa L,mn}(\partial_1\partial_2)^{\kappa-m-n}\square_1^m\square_2^n\Big[(v\partial_1+v\partial_2)^Lf_{\kappa L;m-n}\big(\frac{v\partial_1-v\partial_2}{v\partial_1+v\partial_2}\big)\Big]_0\circ(x_{12}^2)^d$$
where $[\dots]_0$ stands for the harmonic part with respect to
$v\in\RR^{1,3}$, the polynomials $f_{\kappa L;m-n}$ are given by
\eref{fdelta}, and the coefficients $c_{\kappa L;mn}$ solve the recursion
\eref{rec1}, (\ref{rec2}). }

It may be interesting to note that $f_{\kappa L;0}$ are multiples of derivatives of
Legendre polynomials (cf.\ \cite[Eqs.\ 15.2.2, 15.4.4]{AS}):
\bea 
f_{\kappa L;0}(r) =
\frac{2^{\kappa-1}L!}{(L+\kappa)_{\kappa-1}}\cdot 
\partial_r^{\kappa-1}P_{L+\kappa-1}(r). 
\eea
so that, by \eref{Apm}, all functions $f_{\kappa
  L;mn}(r)$ are derivatives of the Legendre polynomials
$P_{L+\kappa-1}(r)$. E.g., for twist 2 ($\kappa=1$), we 
have 
$$e_{1L}(p,q,r)=\big(1+\frac p2(r-1)\partial_r+\frac q2(1+r)\partial_r\big)P_L(r).$$

\medskip

The next task is to relax the assumption $\mu_1=\mu_2=$ scalar, and to
find and solve the analogue of \eref{itwS} in the general case. This
will be necessary in order to compute the contributions from all
insertions of projectors as in \eref{PWE} by successive reduction
according to \eref{reduction}.

For two scalar fields of different dimensions, $d_1\neq d_2$, 
the ansatz $\wh E_\lambda = E_{\kappa L} \circ
(x_{12}^2)^{(d_1+d_2)/2}$ is solved by a scalar polynomial $E_{\kappa L}(\partial_1,\partial_2,v)$, homogeneous of degree $2\kappa+L$ in
$\partial_i$, homogeneous of degree $L$ and harmonic in $v$, as
before, but now satisfying the differential equation 
$$\big(2(\partial_1\nabla_1)\nabla_1-\partial_1\nabla_1^2 + 2(\partial_2\nabla_2)\nabla_2-\partial_2\nabla_2^2
 + (d_1-d_2)(\nabla_1-\nabla_2)\big) E_{\kappa L}(\partial_1,\partial_2,v)=0.
$$
Note that the homogeneity conditions require that $\kappa$ is an
integer, and that in a GCI theory, fields with even twist $2\kappa$
can arise in the OPE only if $d_1-d_2$ is even. One would therefore
have to modify the ansatz when $d_1-d_2$ is odd. 

Similarly, in the chiral case, the ansatz $\wh E_h^{d_1,d_2} = E_h^{d_1,d_2}\circ (x_{12})^{d_1+d_2}$
implies that $E_h^{d_1,d_2}$ is homogeneous of degree $h$ in $\partial_i$ and
satisfies the differential equation 
$$\big(\partial_1\nabla_1^2 + \partial_2\nabla_2^2 +
(d_1-d_2)(\nabla_1-\nabla_2)\big) E_h(\partial_1,\partial_2) =0.$$
This is solved by
\bea\label{chiral-d}
E_h^{d_1,d_2}(\partial_1,\partial_2) = 
\sum_{p+q=h} \frac{(q-d_1+d_2)_p}{p!} \frac{(p+d_1-d_2)_q}{q!}\; \partial_1^p(-\partial_2)^q.\qquad
\eea

\section{Application: Test of positivity of a 6-point structure}
\label{Exo}
\setcounter{equation}{0}

Recall the positivity problem for the exotic scalar 6-point structures
addressed in the introduction.  
We consider here only the simplest example of such a structure, which
has double poles and is consistent with the constraints due to the
requirement that the OPE in both the first and last pair of fields
starts with twist 2 \cite{NRT06}. 
More general double pole structures have been classified in \cite{B09}.

\medskip

In \cite{NRT06}, the leading part of this structure was displayed. 
In \cite{NRT07}, its ``tetraharmonic completion'' (i.e., the
biharmonic completion in both pairs of variables $x_1,x_2$ 
and $x_5,x_6$) was presented in terms of a transcendental function $g(s,t)$. The
tetraharmonic completion is precisely the twist-2 part in both
channels. Unfortunately, however, due to a wrong resummation factor,
this function $g(s,t)$ was incorrectly computed in \cite{NRT07}. We
shall display the correct 
function below.  

The leading part of the exotic structure for four scalar fields
$\phi_1,\phi_2,\phi_5,\phi_6$ of dimension $d$ and two scalar fields
$\phi_3,\phi_4$ of dimension $d'$ is given by 
\bea\label{E6}
E(x_1,\cdots,x_6) =   \frac{\left(x^2_{15}x^2_{26}x^2_{34} -
    2x^2_{15}x^2_{23}x^2_{46}  - 2x^2_{15}x^2_{24}x^2_{36}\right)_{[1,2][5,6]}}
  {(x^2_{12})^{d-1}\cdot x^2_{13}x^2_{14}x^2_{23}x^2_{24}\cdot
  (x^2_{34})^{d'-3}\cdot x^2_{35}x^2_{45}x^2_{36}x^2_{46}\cdot
  (x^2_{56})^{d-1}}\;,\qquad
\eea
where $(\cdot)_{[k,l]}$ stands for antisymmetrization.
Without loss of generality, we choose $d=d'=3$. For comparison, we
also introduce the following 6-point structure with the same
symmetries as $E$, but which has no double poles and appears as part of the 6-point
function of six cubic Wick products of a complex massless scalar free field:
$$B(x_1,\cdots,x_6) = \frac1{(x_{12}^2)^2}
\cdot
\left(\frac1{x^2_{14}x^2_{23}}\right)_{[1,2]}
\cdot
\frac1{x^2_{34}}
\cdot
\left(\frac1{x^2_{36}x^2_{45}}\right)_{[5,6]}
\cdot
\frac1{(x^2_{56})^{2}}
.$$
The structure $B$ is separately biharmonic in both the $1$-$2$ and $5$-$6$
channels. It turns out that the tetraharmonic completion $H$ of
$B-\frac12E$ can be written more compactly than that of $E$ given in \cite{NRT07}, namely 
\bea\label{tetra}
H(x_1,\cdots,x_6)=\big(B-\frac E2\big)\cdot g(s,t)g(s',t'),
\eea
where $s=\frac{x_{12}^2x_{34}^2}{x_{13}^2x_{24}^2}$,
$t=\frac{x_{14}^2x_{23}^2}{x_{13}^2x_{24}^2}$, and
$s'=\frac{x_{34}^2x_{56}^2}{x_{35}^2x_{46}^2}$,
$t'=\frac{x_{36}^2x_{45}^2}{x_{35}^2x_{46}^2}$. 
The condition of biharmonicity amounts
to the differential equation \cite{NRT07}
$$\big[(1-t\partial_t)(1+t\partial_t+s\partial_s) - 
\big((1-t\partial_t)+t(2+t\partial_t+s\partial_s)\big)\partial_s\big]\, g=0
$$
for the function $g(s,t)$. The expansion in a power series in $s$,
$g(s,t)=\sum_n\frac {s^n}{n!}g_n(t)$, gives the recursion
$(1+(n+1)t-t(1-t)\partial_t)g_{n} = 
(1-t\partial_t)(n+t\partial_t)g_{n-1}$ with $g_0(t)=1$. This can be
solved in terms of hypergeometric functions, giving 
\bea\label{gst}
g(s,t)=\sum_n s^n \, \frac{n!(n+1)!}{(2n+1)!}\cdot{}_2F_1(n,n+1;2n+2;1-t).
\eea
The sum can be performed when the integral representation of the
hypergeometric functions \cite[Eq.\ 15.3.1]{AS} is inserted, and $s,t$
are expressed in terms of the ``chiral variables'' $u_\pm$ such that
$s=u_+u_-$ and $t=(1-u_+)(1-u_-)$. Then 
\bea \notag
g(s,t)&=& \sum_n (n+1) s^n\int_0^1 dx
x^{n}(1-x)^{n}(1-(1-t)x)^{-n} \\ \notag
&=& \int_0^1 dx
\Big[\frac{1-(u_++u_--u_+u_-)x}{(1-u_+x)(1-u_-x)}\Big]^2  \\ \notag
 &=& 
1+2u_+u_-\cdot\frac{(1-u_+)(1-u_-)\cdot\log\frac{1-u_+}{1-u_-} +
  u_+-u_--\frac12u_+^2+\frac12u_-^2 }{(u_+-u_-)^3}
\\ \notag  &=&  1 + \sum_{a,b\geq 1} \frac{2ab}{(a+b)((a+b)^2-1)}
u_+^au_-^b  
\\ \label{guv}
 &=& \frac{(1-u_+)(1-u_-)}{u_+-u_-}\cdot \!\!\sum_{a,b\geq 0,a+b>0} 
\frac{a-b}{a+b}\, u_+^au_-^b  \,.
\eea
(In the first line, we corrected a wrong factor of $n!$, whose
presence in \cite{NRT07} and spoiled the subsequent expressions.)

\medskip

Because the twist-2 part is obtained by inserting projections, it must
separately satisfy Wightman positivity. Of course, we would like to
apply the twist-2 intertwiners $E_{1 L}$ of Sect.\ \ref{4Dint} in both channels, so that the
issue reduces to the positivity of tensor-scalar-scalar-tensor 4-point
functions. Applying successively the unknown intertwiners for the
resulting tensor-scalar channels, the problem would be reduced to the
positivity of the resulting 2-point function, i.e., to the positivity
of the numerical amplitude.  

Since we know the intertwiners $E_{1 L}$, the first step can in
principle be done. Notice that it is sufficient to act on the leading
part, because it differs from the twist-2 part by contributions of
higher twist, that are annihilated by $E_{1 L}$. Notice also that $B$
has the form of a product of two 4-point functions in 
the variables $x_1,x_2,x_3,x_4$ and in the variables
$x_3,x_4,x_5,x_6$. Therefore, the application of the intertwining
differential operators in the $1$-$2$ channel and in the $5$-$6$
channel also factorizes. The same, however, is not true for $E$.

Thus, even the first step at present seems to be too involved to be
carried out in practice. The second step is at present not possible
because we have not yet determined the tensor-scalar intertwiners. 

For this reason, we decided to perform only a weaker test of positivity.
Namely, we restrict the twist-2 structure to 2D, by setting two
spatial coordinates to 0. Since this essentially amounts to a smaller
class of test functions, Wightman positivity must still be preserved;
but notice that 2D positivity after the restriction is necessary but not
sufficient to ensure positivity in 4D.  

The intertwining operators in 2D are at our disposal
\eref{chiral}, and we have computed all coefficients (see below).  
It turns out that the partial wave amplitudes of the restricted exotic
twist-2 structure $B-\frac12E$ differ from those of the non-exotic structure
$B$ only by certain signs. This means that $E$ has the same
partial wave amplitudes as $4B$, except that some of them are
absent. 

The non-exotic structure $B$ may itself be indefinite, but we know that it
occurs in a free-field model, and therefore can be dominated by other
positive free-field structures, because free fields are manifestly
positive. This seems to indicate that the restricted exotic structure
as well can be dominated by positive free-field structures. Thus
positivity at the 6-point level alone would not forbid the appearance
of this structure as part of a 6-point correlation function. 

\medskip

Let us indicate some details of the actual computations.

Upon restriction to 2D, $u_+$ and $u_-$ turn into the
chiral cross ratios $u=\frac{x_{12}x_{34}}{x_{13}x_{24}}$. Moreover,
the function $B-\frac12E$ drastically simplifies:
$$B-\frac12E\stackrel{2D}=\frac{1}{(x_{12}^2)^2x_{13}^2x_{24}^2\cdot x_{34}^2\cdot
  x_{35}^2x_{46}^2(x_{56}^2)^2}\cdot \frac{(u_+-u_-)}{(1-u_+)(1-u_-)} \cdot 
\frac{(u'_+-u'_-)}{(1-u'_+)(1-u'_-)}.$$
After multiplication with $g(s,t)g(s',t')$, using \eref{guv}, we
have  
$$
H\stackrel{2D}=\frac{1}{(x_{12}^2)^2x_{13}^2x_{24}^2\cdot x_{34}^2\cdot
  x_{35}^2x_{46}^2(x_{56}^2)^2}\cdot \sum_{a,b\geq 0,a+b>0} 
\frac{a-b}{a+b}\, u_+^au_-^b \cdot \sum_{a,b\geq 0,a+b>0} 
\frac{a-b}{a+b}\, u_+^{\prime\, a}u_-^{\prime\, b}.
$$
For the non-exotic structure $B$, one has instead
$$B\stackrel{2D}=\frac{1}{(x_{12}^2)^2x_{13}^2x_{24}^2\cdot x_{34}^2\cdot
  x_{35}^2x_{46}^2(x_{56}^2)^2}\cdot \sum_{a,b\geq0,a+b>0} u_+^au_-^b\cdot \sum_{a,b\geq0,a+b>0} u_+^{\prime\, a}u_-^{\prime\, b}.$$
Because the sums factorize, the evaluations of the chiral intertwining
differential operators $\iota\circ E_{h_\pm}(\partial_k,\partial_l)\circ (x_{kl,\pm})^d$ in
the $1$-$2$ channel ($k,l=1,2$) and in the $5$-$6$ channel, with $E_h$ given by \eref{chiral},
completely decouple. Actually, because all
structures of interest are of order $x_{kl}^{\geq1-d}$, and therefore
only chiral dimensions $h\geq 1$ will occur, we found it more
efficient to work with chiral intertwining operators $\iota\circ
D_{h_\pm}\circ (x_{kl,\pm})^{d-1}$ where $D_h(\partial_k,\partial_l) =
(\nabla_k-\nabla_l)E_h(\partial_k,\partial_l)$, and adopt a 
normalization different from \eref{chiral}:  
$$D_h(\partial_1,\partial_2) = \frac1{(h-1)!}\sum_{p+q=h-1}\frac{\partial_1^p(-\partial_2)^q}{p!^2q!^2}.$$
Thus, we apply $\iota\circ D_{h_+,h_-}\circ (x_{12}^2)^2 = 
\iota\circ \big[D_{h_+}\circ (x_{12,+})^2\otimes D_{h_-}\circ
(x_{12,-})^2\big]$.  We find 
$$\iota\circ D_h\Big[\frac{1}{x_{13}x_{24}}\, u^a\Big] 
  = x_{34}^a\cdot\iota\circ D_h\, 
  \Big[\frac{x_{12}^a}{(x_{13}x_{24})^{a+1}}\Big]    
= (-1)^{h-1}c_{a,h}\cdot\frac{x_{34}^{h-1}}{(x-x_3)^h(x-x_4)^h}$$
where $c_{a,h}=\frac{(h)_a(1-h)_a}{a!^2}$.
Multiplying the two chiral factors and performing the sum over $a$ and
$b$ gives for the structure $B$
$$\iota\circ D_{h_+,h_-} \Big[\frac{1}{x_{13}^2x_{24}^2} \sum_{a,b\geq 0,a+b>0}  u_+^au_-^b \Big]= C_B(h_+,h_-)\cdot
\frac{x_{34,+}^{h_+-1}}{(x-x_3)_+^{h_+}(x-x_4)_+^{h_+}}\Big[+\to-\Big]$$
where, by virtue of $F(z):=\sum_a c_{a,h}z^a
  =P_{h-1}(1-2z)$ and $P_L(-1)=(-1)^L$,  
\bea C_B(h_+,h_-) = 
2\chi_{\rm odd}(h), \eea
where $\chi_{\rm odd}(h)=1$ if the helicity $h=h_+-h_-$ is odd, and
zero otherwise. To perform the corresponding computation for the sum
weighted with $\frac{a-b}{a+b}$, as in the structure $H$, one may for
$b>0$ put $G_b(z)=\sum_a \frac{a-b}{a+b} c_{a,h}z^a$, solve the equation
$zG'+bG=zF'-bF$ by $G(z)=F(z) -2bz^{-b}\int_0^z t^{b-1}F(t)dt$, and
use the orthogonality of the Legendre polynomials to conclude
$G(1)=F(1)=(-1)^{h-1}$ if $h_+>h_-$. One finds 
\bea C_H(h_+,h_-) = \mathrm{sign}(h)\cdot 
2\chi_{\rm odd}(h). \eea
The same factors arise in the $5$-$6$ channels. Thus, when the 6-point
structures $B$ and $H$ are reduced in both channels by means of $(\iota_x \circ
D_{h_+,h_-}\circ (x_{12}^2)^2)\otimes (\iota_{x'} \circ
D_{h'_+,h'_-}\circ (x_{56}^2)^2)$, the result is always a
multiple of the same 4-point function
$$W_{h_+,h_-;h_+',h_-'}(x,x_3,x_4,x')=\frac{x_{34,+}^{h_++h_+'-3}}{(x-x_3)_+^{h_+}(x-x_4)_+^{h_+}(x_3-x')_+^{h_+'}(x_4-x')_+^{h_+'}}\times\Big[+\to-\Big]. $$
The respective coefficients for the structures $B$ and $H$ are
\bea C_B(h_+,h_-)C_B(h'_+,h'_-)&=&
4\chi_{\rm odd}(h)\chi_{\rm odd}(h')\, , \notag \\
C_H(h_+,h_-)C_H(h'_+,h'_-)&=&\mathrm{sign}(h)\mathrm{sign}(h')\cdot
4\chi_{\rm odd}(h)\chi_{\rm odd}(h')\, , 
\eea
where $h=h_+-h_-$, $h'=h_+'-h_-'$ are the helicities.

Because $H$ is the twist-2 part of $B-\frac12E$, we conclude that
(after 2D restriction) all partial waves with helicities of
equal sign in the $1$-$2$-channel and in the $5$-$6$-channel, that are
present in $B$, are absent in the twist-2 part of $E$, while those
with helicities of opposite sign arise in the twist-2 part of $E$ with 4
times the coefficient in $B$. 

\medskip

It remains to perform the partial wave expansion of the 4-point
functions $W_{h_+,h_-;h_+',h_-'}$. Here one may use standard methods,
e.g., \cite{RS,DO,NRT05}. Namely, 
$$W_{h_+,h_-;h_+',h_-'}(x,x_3,x_4,x') =
\sum_{k_+,k_-}B^{k_+,k_-}_{h_+,h_-;h_+',h_-'} \cdot
W^{k_+,k_-}_{h_+,h_-;h_+',h_-'}(x,x_3,x_4,x'),$$
where $W^{k_+,k_-}$ is the partial wave for the insertion of a
projection on the representation with scaling dimensions
$(k_+,k_-)$. It turns out that only $k_\pm\in\frac32+\NN_0$
contribute. 
Because of chiral factorization of $W_{h_+,h_-;h_+',h_-'}$, one has
$B^{k_+,k_-}_{h_+,h_-;h_+',h_-'}=B^{k_+}_{h_+,;h_+'}B^{k_-}_{h_-,;h_-'}$,
where the chiral coefficients are determined by the expansion
$$1 = \sum_{k=\frac32+n}B^{k}_{h,h'}\cdot
u^{n}\,{}_2F_1(n+h,n+h';2n+3;u).$$
The problem of Wightman positivity of the (2D-restricted) structures $B$
and $H$ has now been reduced to the positivity of linear combinations
of matrices of the form  
$$P_\pm P_{\rm odd}\big[B^{k_+}\otimes B^{k_-}\big]P_{\rm odd}P_\pm,$$
where $P_{\rm odd}$ and $P_\pm$ are the projections on the odd resp.\
positive or negative helicities. 

To be admissible in a QFT, the exotic structure does not need to be
separately positive, but must only be dominated by other, non-exotic
structures that contribute to a full 6-point function. Thus, if
positivity should fail for $H$ (it certainly does for the twist-2 part
of $E$ because in this case all diagonal matrix elements vanish), one
would have to establish a bound for the negative part of the above
matrices by positive matrices of partial wave amplitudes arising from
other structures.  

We have not completed this analysis yet.

\medskip

To conclude: the tools are available to test Wightman positivity of
6-point correlation functions. If a 6-point function involving the
exotic structure \eref{E6} passes the test, then it could be a candidate
for a nontrivial 4D conformal QFT.

\begin{acknowledgement}
KHR is grateful for helpful discussions with N.M.~Nikolov and I.~Todorov,
and also with Ch. Mishra (IISER, Kolkata) in an early stage of this work.
Supported in part by the German  Research Foundation
(Deutsche Forschungsgemeinschaft (DFG)) through the Institutional
Strategy of the University of  G\"ottingen.
\end{acknowledgement}

\end{document}